\documentclass[aps,pra,groupedaddress,twocolumn]{revtex4}   
\usepackage{graphicx}
\usepackage{float}
\usepackage{epsfig}
\usepackage{hyperref}
\usepackage{amssymb, amsmath, amsthm}
\usepackage{subcaption}
\usepackage{url}
\usepackage{xcolor}
 
\newcommand{\norm}[1]{\left\Vert#1\right\Vert}
\newcommand{\abs}[1]{\left\vert#1\right\vert}
\newcommand{\set}[1]{\left\{#1\right\}}

\newcommand{\trace}[1]{\mbox{tr}\left( #1 \right)}

\newcommand{\refe}[1]{(\ref{#1})}

\usepackage{comment}

\newcommand{\IFLP}{ Instituto de F\'isica La Plata (CONICET-UNLP), La Plata,
Buenos Aires, Argentina}

\newcommand{\ICC}{
Instituto de Ciencias de la Computaci\'on, CONICET - Universidad de Buenos Aires, Buenos Aires, Argentina}

\newcommand{\CITECCA}{Centro Interdisciplinario de Telecomunicaciones Electr\'onica, Computaci\'on y Ciencia Aplicada (CONICET), Universidad Nacional de Rio Negro, Bariloche, Rio Negro, Argentina}

\begin{document}

\title{Parametrized Variational Quantum Tomography}

\author{V. A. Penas}
\email[]{vapenas@unrn.edu.ar}
\affiliation{\CITECCA}

\author{M. Losada}
\affiliation{\ICC}

\author{D. Tielas}
\affiliation{\IFLP}

\author{F. Holik}
\affiliation{\IFLP}

\date{\today}

\begin{abstract}

Quantum state tomography provides a fundamental framework for reconstructing quantum states. 
When the measurement data are not informationally complete, the observed statistics admit multiple compatible density matrices, making the reconstruction problem inherently underdetermined and calling for the selection of a meaningful estimator. Two well-established approaches to address this ambiguity are Maximum Entropy (MaxEnt) and Variational Quantum Tomography (VQT).

A variant of VQT, named VQT$_\infty$, has been introduced to reproduce MaxEnt-like solutions.
In this work, we generalize this approach by introducing a parametrized cost function that interpolates between the 1-norm and the infinity norm, thereby unifying VQT and VQT$_\infty$ within a single framework. By tuning the associated hyperparameters, the proposed method enables controlled exploration of the set of compatible density matrices. We show that this interplay yields reconstructed states with  higher fidelity to the MaxEnt solution than those obtained via VQT$_\infty$ while preserving computational tractability. 
\end{abstract}

\maketitle

\section{Introduction}\label{new_intro}

Quantum State Tomography aims to reconstruct the density matrix of a quantum system from measurement outcomes \cite{ncbook, clavor2011, maciel2012}. Its main limitation lies in scalability: the number of measurements required for an informationally complete reconstruction grows exponentially with the size of the system \cite{MauroDAriano2003,Paris2004QuantumStateEstimation, Cramer2010}, rendering full tomography impractical for systems composed of many qubits.
This challenge motivates alternative approaches that circumvent exponential scaling, including incomplete-data methods for approximate reconstruction \cite{gross2010,Flammia2012}, structure-exploiting approaches \cite{Cramer2010,avella2025efficientstateestimationquantum}, property-targeted estimation \cite{Bendersky2008,Schmiegelow2011}, and compressed representations enabling the prediction of many observables without full reconstruction \cite{Aaronson2020,Huang2020}. In this work, we adopt the incomplete-data paradigm, focusing on quantum state estimation from partial measurements.

When measurement data are incomplete, the quantum state is underdetermined, since multiple density matrices may reproduce the observed outcomes. In such situations, the Maximum Entropy (MaxEnt) principle provides a natural inference rule by selecting the state that maximizes the von Neumann entropy subject to the constraints imposed by the available data. The resulting estimator is unique and corresponds to the least biased state compatible with the measurements.

In practice, however, implementing this principle generally requires solving computationally demanding nonlinear optimization problems. In some cases, there are strategies to reduce computational complexity, such as considering  state symmetries \cite{PhysRevLett.105.250403,CORTE2023,Losada2019,Tielas2022,Holik2025}.

An alternative strategy introduced by Maciel et al. \cite{maciel2011}, Variational Quantum Tomography (VQT), addresses this difficulty by formulating the reconstruction problem as a linear semidefinite program (SDP) \cite{klerkbook2002,boyd2004}. VQT provides a computationally tractable method for reconstructing quantum states from incomplete and noisy data. A key feature of this formulation is that it yields a physically consistent density matrix without overestimating purity or entanglement.

A modified version of VQT, called VQT$_{\infty}$, was later proposed by Gonçalves et al. \cite{Goncalves2013}. This formulation yields estimates similar to the maximum entropy solution by distributing the unmeasured probabilities as uniformly as possible. At the same time, it retains the computational advantage of being efficiently solvable via semidefinite programming.

Despite these appealing features, the analysis presented in  \cite{Goncalves2013} exhibits certain limitations. In particular, the numerical study focuses exclusively on the average fidelity obtained from randomly generated states, without addressing the full distribution of fidelities or directly evaluating the fidelity between the MaxEnt and VQT$_{\infty}$ estimated states. 
Consequently, how the reconstruction performance behaves beyond average values remains an open question.

In this work, in first place, we examine the relationship between VQT$_\infty$ and the MaxEnt approach with greater detail.  Rather than focusing solely on average estimates we analyze and compare the distributions of fidelities of reconstructed states. This allows us to assess whether the equivalence between VQT$_\infty$ and MaxEnt extends beyond average values and holds at the level of statistical behavior.

In doing so, a natural question arises as to whether the SDP defined with VQT$_{\infty}$ offers any special advantage in a non-eigenbasis case or can be modified to accommodate for better results against states estimated using MaxEnt. We thus introduced a parametrized version of VQT where the cost function includes both 1-norm and infinity norm of VQT and VQT$_{\infty}$, respectively.
By systematically comparing several instances of random states of up to five qubits, we obtain a more detailed characterization of the relationship between these methodologies.

This works is organized as follows. In Section \ref{sec:QST} we provide a short overview of the MaxEnt, VQT and VQT$_\infty$ approaches. In Section \ref{sec:PVQT} we introduced parametrized variational quantum tomography (PVQT).  Next, in Section \ref{sec:NumericalSection}, we present numerical results for three-, four-, and five-qubit states. Finally, in Section \ref{concl} we draw our conclusions.

\section{Quantum state tomography}\label{sec:QST}

Quantum state tomography aims to reconstruct the density matrix of a quantum system from measurement outcomes. When the available data are not informationally complete, the observed statistics admit multiple compatible density matrices, rendering the reconstruction problem inherently underdetermined and requiring the selection of a meaningful estimator. This task is commonly addressed by approaches such as Maximum Entropy (MaxEnt), Variational Quantum Tomography (VQT), and its variant VQT$_\infty$.

In this work, the available data consist of measurements corresponding to a subset of a POVM. Let $\set{E_i}_{i=1}^N$ be a POVM, where the subset  $\set{E_i}_{i=1}^K$ is measured, while the subset  $\set{E_i}_{i=K+1}^N$ remain unmeasured.
We define 
the vector of unmeasured probabilities
$\vec{u}$ as follows
\begin{equation}
\label{vec_u}
\vec{u} = \left( \mathrm{tr}(\rho E_{K+1}), \ldots, \mathrm{tr}(\rho E_{N}) \right).
\end{equation}

In the following, we briefly review MaxEnt, VQT, and VQT$_\infty$.

\subsection{Maximum Entropy estimation}\label{maxentsec}

The application of the Maximum Entropy principle to quantum state tomography with incomplete data was introduced by Buzek et al. \cite{buzek1999}. This approach consists in estimating the density matrix $\rho$ by maximizing the von Neumann entropy,
\begin{equation}
S(\rho) = -\trace{\rho \ln \rho},
\end{equation}
subject to consistency with the available measurement data. 

In this scenario the MaxEnt principle leads to the following optimization problem:
\begin{equation}
\label{maxent}
\begin{aligned}
\underset{\rho}{\text{maximize}} \quad & -\trace{\rho \ln \rho} \\
\text{subject to} \quad 
& \trace{E_i \rho} = f_i, \qquad (i = 1, \ldots  , K)  \\
& \trace{\rho} = 1, \\
& \rho \succeq 0 .
\end{aligned}
\end{equation}

\noindent
where $f_i$'s are the measured frequencies. Assuming a full-rank solution \(\rho \succ 0\), an explicit solution can be obtained:
\begin{equation}
\rho_{\mathrm{ME}} = \frac{1}{\mathcal{N}} e^{ - \sum_{i=1}^{K} \lambda_i E_i },
\label{mee}
\end{equation}
where $\lambda_i$ are the Lagrange multipliers associated to the equality constraints, and $\mathcal{N} = \trace{ \exp  \left(- \sum_{i=1}^{K} \lambda_i E_i\right)}$ is a normalization constant.

The Lagrange multipliers can be determined by solving the non-linear system of equations
\begin{equation}
\trace{E_i \rho_{ME}} \ = f_i,  \ \qquad i = 1, \ldots , K,
\end{equation}
that is,
\begin{equation}
\label{eqmaxent}
\trace{E_i e^{ -\sum_{i=1}^{K}  \lambda_{i'} E_{i'}}} = {\cal N} f_{i},   \ \qquad i = 1, \ldots , K.
\end{equation}
In general, for noisy data, we solve the following non-linear least-squares problem \cite{Goncalves2013}
\begin{equation}\label{eq:nlsq}
\underset{\lambda_i}{\text{minimize}} \ \ \sum_{i'=1}^{K}  \left[\trace{E_{i'} e^{-\sum_{i=1}^{K}  \lambda_i E_i}} - {\cal N} f_{i'} \right]^2,
\end{equation}
instead of the non-linear equations \refe{eqmaxent}.

\subsection{Variational Quantum Tomography}

Variational Quantum Tomography (VQT) was introduced by Maciel et al. \cite{maciel2011} as a computationally efficient method for reconstructing quantum states from incomplete and noisy measurement data. The approach formulates the estimation problem as a linear semidefinite program, ensuring a physically valid density matrix while allowing efficient numerical optimization.

The VQT formulation reads as follows
\begin{equation}
\begin{aligned}
\underset{{\rho, \ \Delta_i}}{\text{minimize}} \  \ & \sum_{i=1}^{K} \Delta_i + \sum_{i=K+1}^{N}  \trace{E_j \rho} \\
\text{subject to} \  \ & \abs{\trace{E_i \rho} - f_i} \le \Delta_i f_i~~~~   (i = 1, \ldots , K), \\
& \Delta_i \geq 0, & \ \\
& \trace{ \rho} \ = 1, & \  \\
& \rho \ \succeq 0,  \\
\end{aligned}
\label{vqt}
\end{equation}
where $\set{\Delta_i}$ are tolerances.

In the case of an ideal, noise-free experiment and, the estimated density matrix $\rho$ must  be consistent with the measured data, namely,
\begin{equation}
\label{linear_eq}
\trace{E_i \rho} = f_i,  \qquad i = 1, \ldots , K.
\end{equation}
To account for experimental noise, the equality constraints are relaxed to: 
\begin{equation}
\label{linear_ineq}
\bigl|\trace{E_i \rho} - f_i\bigr| \le \Delta_i f_i, \qquad i = 1, \ldots , K.
\end{equation}
 Minimizing  $\Delta_i$ tolerances motivates the first term in the objective function.
Higher-quality measurements correspond to smaller values of $\Delta_i$, and in the ideal limit of noiseless measurements with infinite statistics, the $\Delta_i$ tend to zero. 

In the incomplete-data scenario, there may exist multiple density matrices that minimize these tolerances while satisfying the remaining constraints. VQT selects the one that minimizes the expectation value of the unmeasured observables. 
This contribution to the cost function can be written in terms of the vector  $\vec{u}$, given in Eq. \eqref{vec_u}, as follows,
\begin{equation}
\begin{aligned}
\underset{{\rho, \ \Delta_i}}{\text{minimize}} \  \ & \sum_{i=1}^{K} \Delta_i + \norm{\vec{u}}_1 \\
\text{subject to} \  \ & \abs{\trace{E_i \rho} - f_i} \le \Delta_i f_i ~~~  (i = 1, \ldots , K), \\
& \Delta_i \geq 0, & \ \\
& \trace{ \rho} \ = 1, & \  \\
& \rho \ \succeq 0,  \\
\end{aligned}
\label{vqt_u}
\end{equation}

One of the main advantages of VQT is that the formulation in Eq. \eqref{vqt} yields a linear semidefinite program (SDP)~\cite{klerkbook2002,boyd2004,helm1996,nemtodd2008}.

\subsection{VQT$_{\infty}$}

In \cite{Goncalves2013}, a modified version of Variational Quantum Tomography, denoted VQT$_{\infty}$, was introduced. The central idea is to replace the
$1$-norm of the cost function with the infinity norm, i.e., $\norm{\vec{u}}_1$ with $\norm{\vec{u}}_{\infty} = \max_{i = K+1, \ldots, N} \ \trace{E_i \rho}$. 

This results in the following optimization problem:
\begin{equation}
\begin{aligned}
\underset{\rho, \ \Delta_i}{\text{minimize}} \  \ & \sum_{i = 1} ^K \Delta_i + \norm{\vec{u}}_{\infty} \\ 
\text{subject to} \  \ & \abs{\trace{E_i \rho} - f_i} \le \Delta_i f_i ~~~   (i = 1, \ldots , K), \\
& \Delta_i \ \geq 0, & \  \\
& \trace{ \rho} \ = 1, & \  \\
& \rho \ \succeq 0. & \ \\
\end{aligned}
\label{vqtinf}
\end{equation}

This modified VQT is motivated by the fact that, in the case of eigenbasis measurements (i.e., measurements in the eigenbasis of $\rho$) and assuming $\Delta_i=0$, the standard VQT objective function contains no constraint forcing the unmeasured coefficients to agree with those of the MaxEnt solution. 
By contrast, in this case, the solution of VQT$_\infty$ coincides with the MaxEnt solution.

Also, the authors argue that despite the agreement with MaxEnt in the eigenbasis case, the proposed modification does not turn the problem \refe{vqtinf} harder than \refe{vqt}. That is, it can be shown that \refe{vqtinf} is equivalent to the following linear SDP problem:

\begin{equation}
\begin{aligned}
\underset{\rho, \ \Delta_i,\ \delta}{\text{minimize}} \  \ & \sum_{i = 1} ^K \Delta_i + \delta \\ 
\text{subject to} \  \ & \abs{\trace{E_i \rho} - f_i} \le \Delta_i f_i ~~~ (i = 1, \ldots, K),  \\
\   \ \ & \trace{E_i \rho} \le \delta ~~~~~~~~~~~~~~~( i = K+1 , \ldots, N), \\
& \Delta_i \ \geq 0, ~ \delta \ \geq 0,  & \  \\
& \trace{ \rho} \ = 1, & \  \\
& \rho \ \succeq 0. & \ \\
\end{aligned}
\label{vqtinf0}
\end{equation}
where $\delta$ is an auxiliary variable.  

This modification establishes an equivalence between the solution of the VQT and that of MaxEnt, at least in the eigenbasis case, as both methods distribute the unmeasured probabilities as uniformly as possible. 
Furthermore, it was shown numerically in \cite{Goncalves2013} that MaxEnt and VQT$_{\infty}$ yield estimated states at similar average trace distance from the target state when measurements are performed in a SIC-POVM.

\section{Parameterized Variational Quantum Tomography}\label{sec:PVQT}

In this section we introduce a parametrized variational quantum tomography (PVQT) that recovers both VQT and VQT$_{\infty}$
as limiting cases.

As previously stated, VQT and MaxEnt possess different intrinsic bias regarding the treatment of unmeasured probabilities. VQT$_{\infty}$ fixes this problem by modifying the objective function to match the behaviour of MaxEnt at least when measuring in the eingenbasis. However, in the general case, there is no reason  why the term $||\vec{u}||_1$ cannot contribute non-trivially to the cost function in a way that yields higher-fidelity results than VQT$_\infty$ solutions.
Therefore, we propose a simple modification to the objective function of VQT that combines the 1-norm and the infinity norm.
 
More precisely, we propose the following optimization problem:

\begin{equation}
\begin{aligned}
\underset{\rho, \ \Delta_i}{\text{minimize}} \ \ \ & \sum_{i = 1} ^K \Delta_i +\alpha\norm{\vec{u}}_{1}
+\beta\norm{\vec{u}}_{\infty}   \\ 
\text{subject to} \ \ \ & \abs{\trace{E_i \rho} - f_i} \le \Delta_i f_i ~~~~~  (i = 1, \ldots , K), \\
& \Delta_i \ \geq 0, & \  \\
& \trace{ \rho} \ = 1, & \  \\
& \rho \ \succeq 0. & \ \\
\end{aligned}
\label{vqt_adaptive_original}
\end{equation}

It can be shown that \refe{vqt_adaptive_original} is equivalent to the following linear SDP problem:
\begin{equation}
\begin{aligned}
\underset{\rho, \ \Delta_i,\ \delta}{\text{minimize}} \ \ \ & \sum_{i = 1}^K  \Delta_i + \alpha \sum_{i =K+ 1}^N  \mbox{tr}(E_i\rho) + \beta\delta  \\ 
\text{subject to} \ \ \ & \abs{\trace{E_i \rho} - f_i} \le \Delta_i f_i ~~~   (i = 1, \ldots , K),   \\
\  \ \ \ & \trace{E_i \rho} \le \delta  ~~~~~~~~~~~~~~~   (i = K+1, \ldots , N),  \\
& \Delta_i \ \geq 0, ~~ \delta \ \geq 0, & \  \\
& \trace{ \rho} \ = 1, & \  \\
& \rho \ \succeq 0. & \ \\
\end{aligned}
\label{vqt_adaptive}
\end{equation}
$\delta$ is an auxiliary variable just like in the VQT$_\infty$ case. Here, we have introduced two hyperparameters $\alpha$, $\beta$ that interpolates between $1$-norm and infinity norm. We choose $0\le\alpha\le1$ and $0\le\beta\le1$ so the relative weight of these term are on equal footing as the first term of the cost function. Note that we call them hyperparameters because they do not enter as optimization parameters on the SDP. 
Setting $\alpha = 1$, $\beta = 0$ recovers VQT, while $\alpha = 0$, $\beta = 1$ recovers VQT$_{\infty}$.
When both $\alpha$ and $\beta$ are active simultaneously, we expect PVQT to exhibit new behavior beyond a mere interpolation between VQT and VQT$_{\infty}$, as we show in Section \ref{subsec:vqt-hib}.

When activating both $\alpha$ and $\beta$ simultaneously we expect new behavior from the SDP rather than just an interpolation between VQT and VQT$_\infty$ as we show in Section \ref{subsec:vqt-hib}.

\section{Numerical simulations}\label{sec:NumericalSection}

In this section we present a numerical comparison between the tomographic methods VQT$_\infty$, MaxEnt and PVQT. 
We focus primarily on the 
three-qubit case, 
although four- and five-qubit cases exhibit similar behavior
(see Subsection \ref{subsec:4-5q}).
We set the hyperparameter $\alpha=1$ while varying $\beta$. This choice allows us to numerically quantify the non-trivial contribution of $||\vec{u}||_{\infty}$ to the VQT cost function. 

We first present additional numerical results comparing VQT$_{\infty}$ and MaxEnt, not addressed in \cite{Goncalves2013}. We do this by studying  several quantities and analyzing to what extend these methods produce comparable estimated states. Then we present results of the parametrized version of VQT. 

For this purpose, we sample random density matrices according to the Haar measure for various fixed ranks.
\cite{karol2011}. 
The measured observables are chosen as a subset of a SIC-POVM \cite{renes2004,SIC_POVM}. 
From the subset of measured observables and their expectation values, the quantum state is reconstructed using the tomographic methods VQT$_{\infty}$, MaxEnt, and PVQT, as defined in previous sections.

\subsection{VQT$_{\infty}$ and MaxEnt}\label{subsec:vqt-inf-vs-maxent}

In \cite{Goncalves2013}, the comparison between VQT$_{\infty}$ and MaxEnt is restricted to the average trace distance between the reconstructed and target states. In this work, we provide a more comprehensive assessment by also considering the fidelity as a performance metric, reporting histograms of fidelities across random states for varying numbers of measured observables.

To this end, we sample $500$ three-qubit pure states (rank $1$) drawn uniformly at random according to the Haar measure.
We consider subsets of observables of a SIC-POVM and take as measurement outcomes the expectation values of the  observables with respect to the target states (noiseless measurements).

\begin{figure}
\centering
\includegraphics[width=8.8cm]{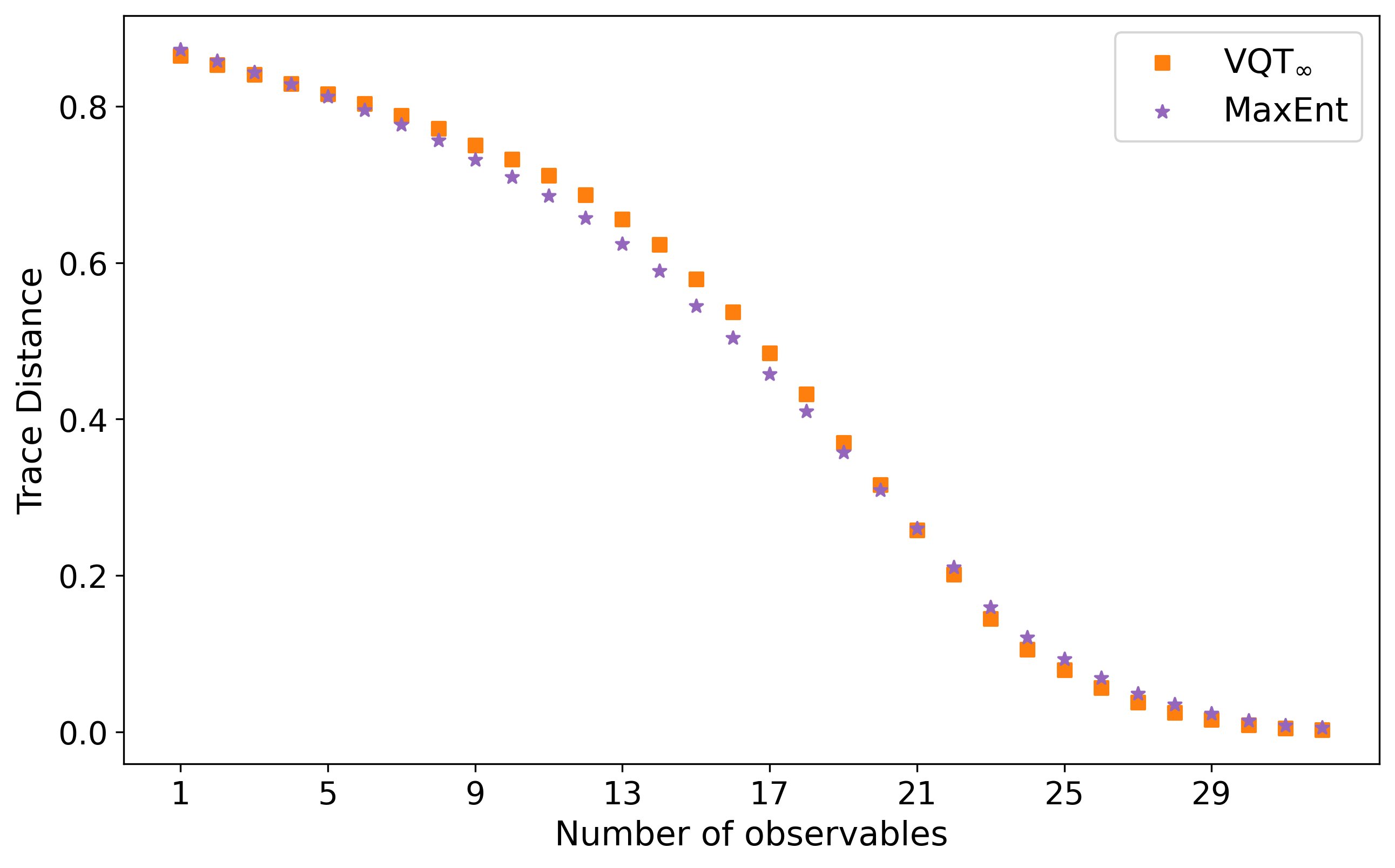}
\caption{Average trace distance  between target and reconstructed states for $500$ random three-qubit  pure states as a function of the number of measured observables, for MaxEnt and VQT$_{\infty}$. The cutoff is set to 32 observables while quorum is 64.}
\label{vqt-inf-maxent-trace-dist-3q}
\end{figure}

\begin{figure}
\centering
\includegraphics[width=8.8cm]{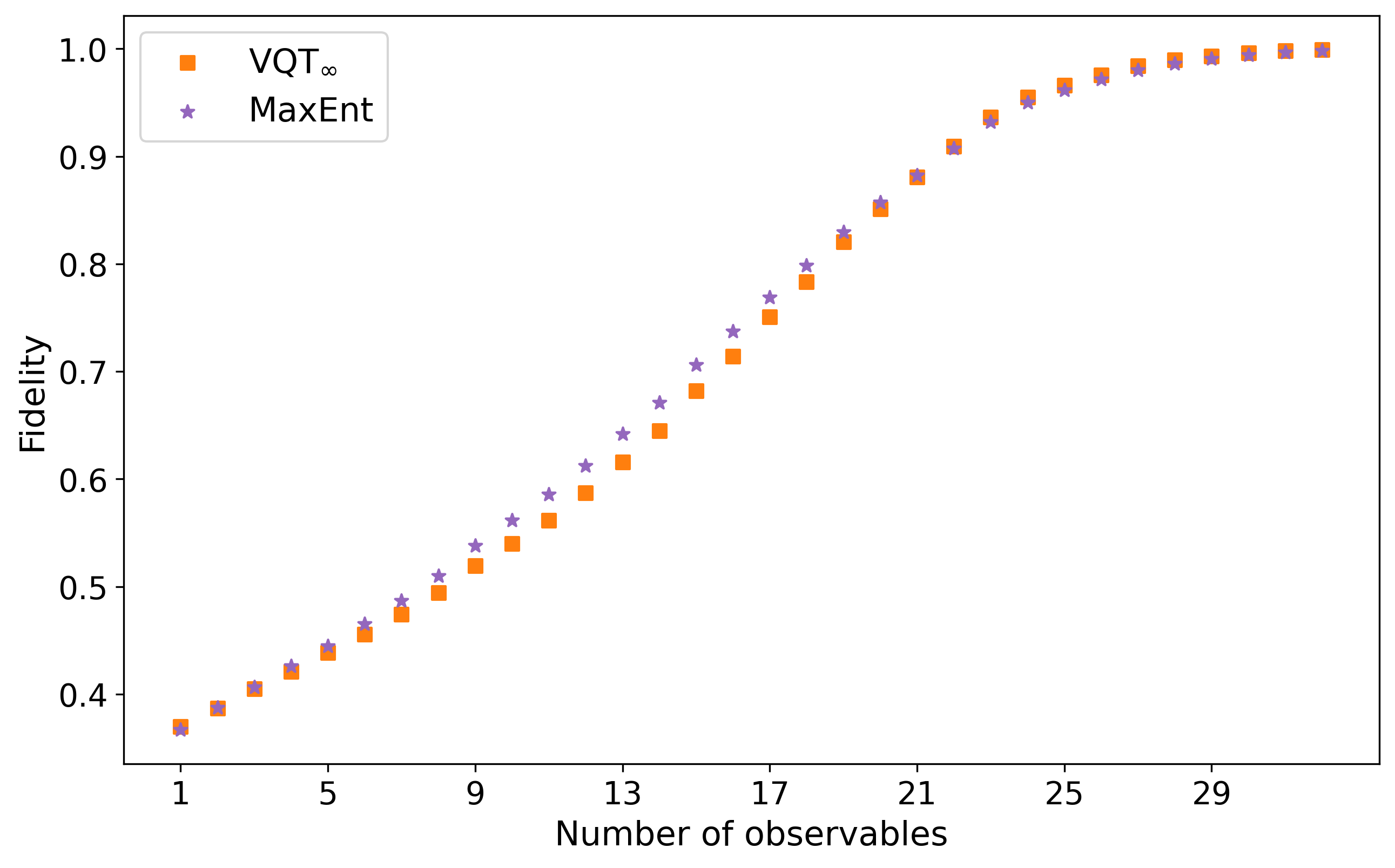}
\caption{
Average fidelity between target and reconstructed states for $500$ random three-qubit  pure states as a function of the number of measured observables, for MaxEnt and VQT$_{\infty}$. The cutoff is set to 32 as in Fig.~\ref{vqt-inf-maxent-trace-dist-3q}.}
\label{vqt-inf-maxent-fid-3q}
\end{figure}

Figs.~\ref{vqt-inf-maxent-trace-dist-3q} and~\ref{vqt-inf-maxent-fid-3q} show the average trace distance and fidelity between target and reconstructed states, respectively, as a function of the number of measured observables, for 500 random three-qubit pure states under the MaxEnt and VQT$_{\infty}$ methods.

\begin{figure}[htbp]
    \centering
    
    \begin{subfigure}{0.455\textwidth}
        \centering
          \caption{}\vspace{-3pt}
        \includegraphics[width=\linewidth]{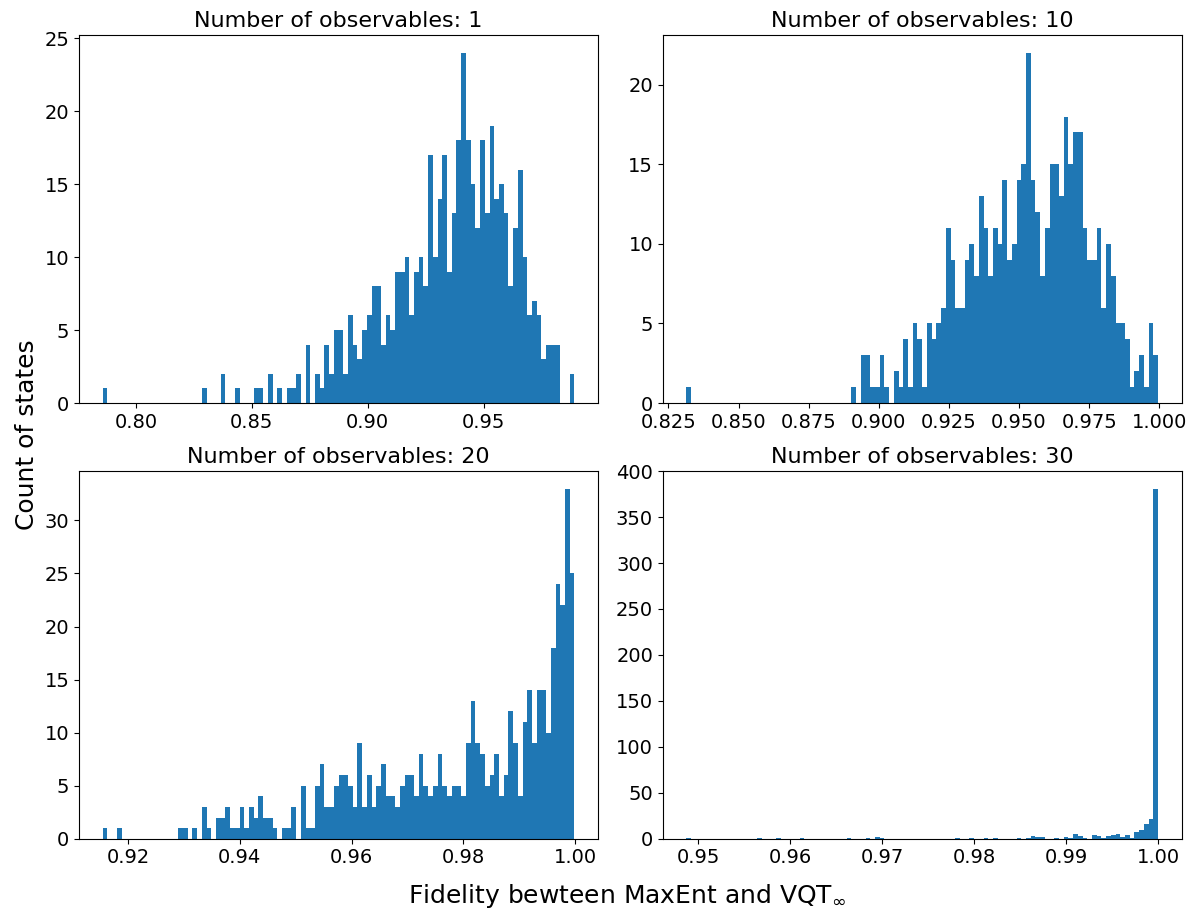}
      %
    \end{subfigure}\vspace{-10pt} 
    \hfill
    \begin{subfigure}{0.43\textwidth}
        \centering
            \caption{}
        \includegraphics[width=\linewidth]{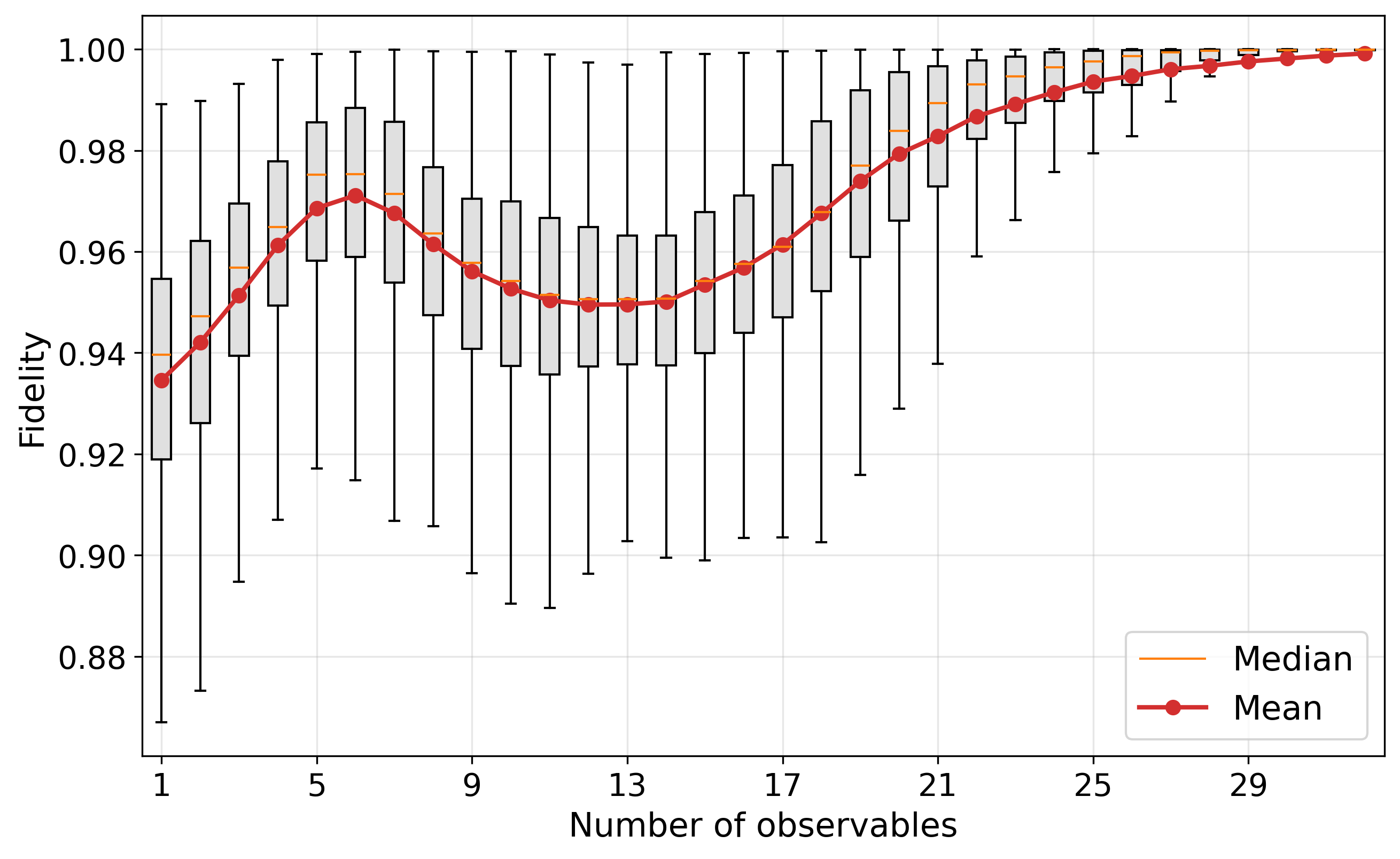}
%
    \end{subfigure}
        \vspace{-9pt}
    \caption{(a) Histograms of fidelities between VQT$_{\infty}$ and MaxEnt  reconstructed states, over $500$ random three-qubit pure states.
    (b) Average fidelity between VQT$_\infty$ and MaxEnt reconstructed states. Error bars represent quartiles. Note that mean (average) and median are close to each other but they are not equal.}
    \label{fig:hist-3q-and-quart-vqt-inf}
\end{figure}

We have used a similar criterium as that of \cite{Goncalves2013}, that is, a method converges if the trace distance to the target state is less than $10^{-4}$. These figures show that both methods yield the same average fidelity relative to the target state. 
Nevertheless, we find that beyond a specific cutoff of measured observables (around 32 observables in this case), the two methods become virtually equivalent as the average fidelity approaches to unity. The number of observables needed to attain this convergence is lower than quorum, meaning that there exists a non trivial regime in which both methods can be used interchangeably. It is important to emphasize, however, that this equivalence is strictly a statistical average; the fidelity distributions are not sharply peaked at one. 
To illustrate this more explicitly, in Fig.~\ref{fig:hist-3q-and-quart-vqt-inf}a, for different numbers of observables, we present histograms of fidelities between VQT$_{\infty}$ and MaxEnt  reconstructed states, over $500$ random three-qubit pure states.

As illustrated by the histograms, the two methods frequently yield similar estimates. The divergence between them diminishes as the number of observables increases, eventually vanishing when both methods converge to the target state. The Maximum Entropy approach provides the least biased estimation compatible with the available data. Therefore, any significant deviation of the VQT$_{\infty}$ reconstruction from the MaxEnt result would introduce an undesirable bias. 
This also means that target-state fidelity alone is an insufficient metric: methods that agree on average with the target may still differ considerably from one another.
To assess this directly, in Fig.~\ref{fig:hist-3q-and-quart-vqt-inf}b we plot the fidelity between the states reconstructed by VQT$_{\infty}$
and MaxEnt.

As seen in the figures, VQT$_{\infty}$ and MaxEnt yield closely matching reconstructions, though not identical ones. The absence of exact equivalence is expected, given that SIC-POVM measurements do not correspond to an eigenbasis, removing any theoretical grounds for the two methods to coincide exactly. Despite this, the reconstructed density matrices are in remarkably good agreement throughout.

\subsection{Parametrized Variational Quantum Tomography}\label{subsec:vqt-hib}
Using the same numerical settings as in the previous subsection, we analyze PVQT, fixing $\alpha = 1$. We present results for $\beta = 0.01$ and $\beta = 0.003$. 
For $\beta = 1$, the three-qubit results are virtually indistinguishable from those of the VQT$_{\infty}$ method, a behavior that persists for a broad range of $\beta$ values below 1. We therefore report results starting from $\beta = 0.01$, where meaningful differences begin to emerge.

\begin{figure}
\centering
\includegraphics[width=8.8cm]{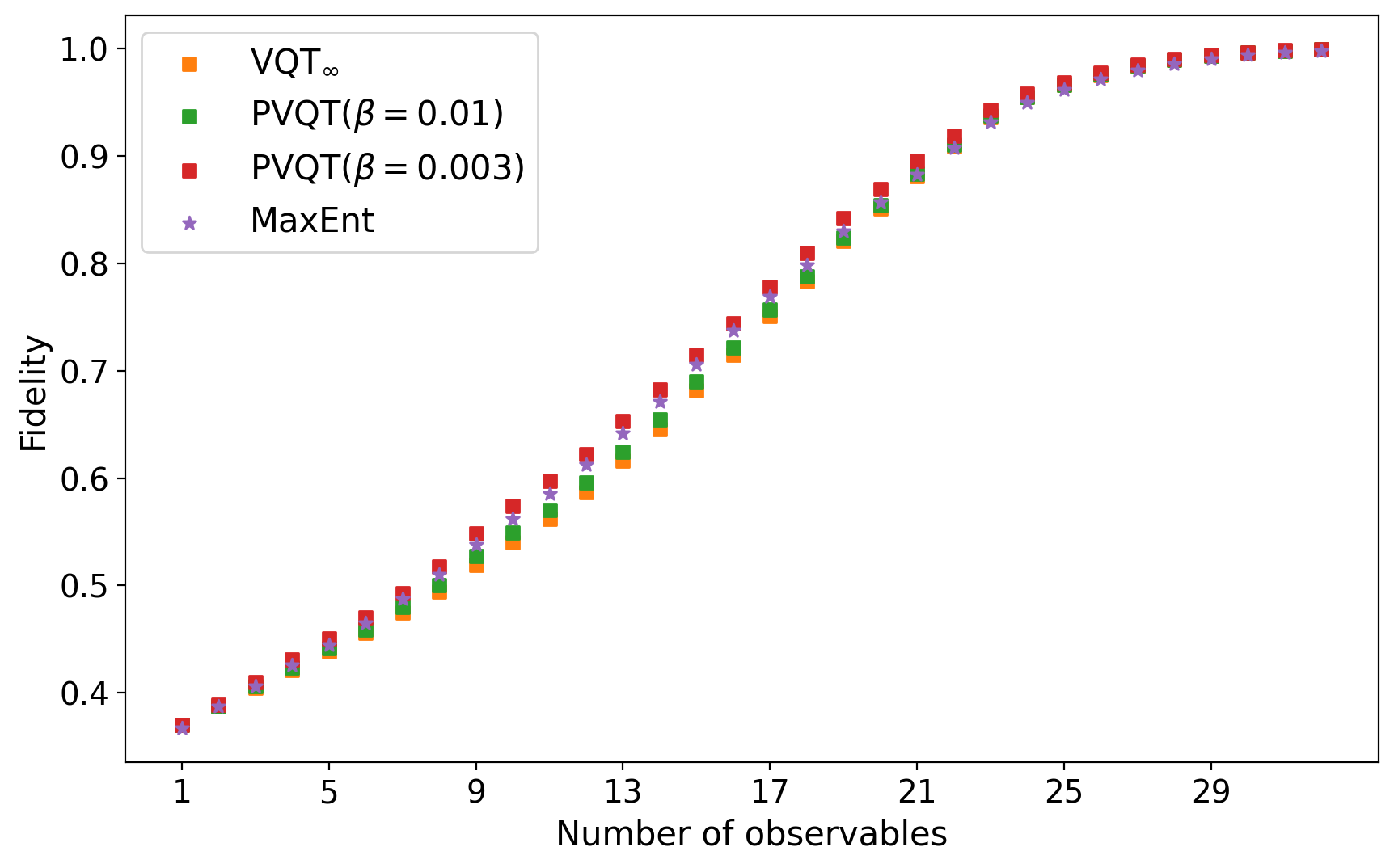}
\caption{Average fidelity between target and reconstructed states for $500$ random three-qubit pure states as a function of the number of measured observables, for MaxEnt, VQT$_{\infty}$, and PVQT with $\beta = 0.01$ and $\beta = 0.003$.}
\label{contra-target-fid-todo-junto-3q}
\end{figure}

We proceed to evaluate MaxEnt and PVQT with respect to the target states, as shown in Fig.~\ref{contra-target-fid-todo-junto-3q}.
We observe that varying parameter $\beta$ affects the average fidelity with respect to the target state. As discussed previously, larger values of $\beta$ approach the behavior of VQT$_{\infty}$. 

\begin{figure}[htbp]
    \centering
    
    \begin{subfigure}{0.455\textwidth}
        \centering
          \caption{}\vspace{-3pt}
        \includegraphics[width=\linewidth]{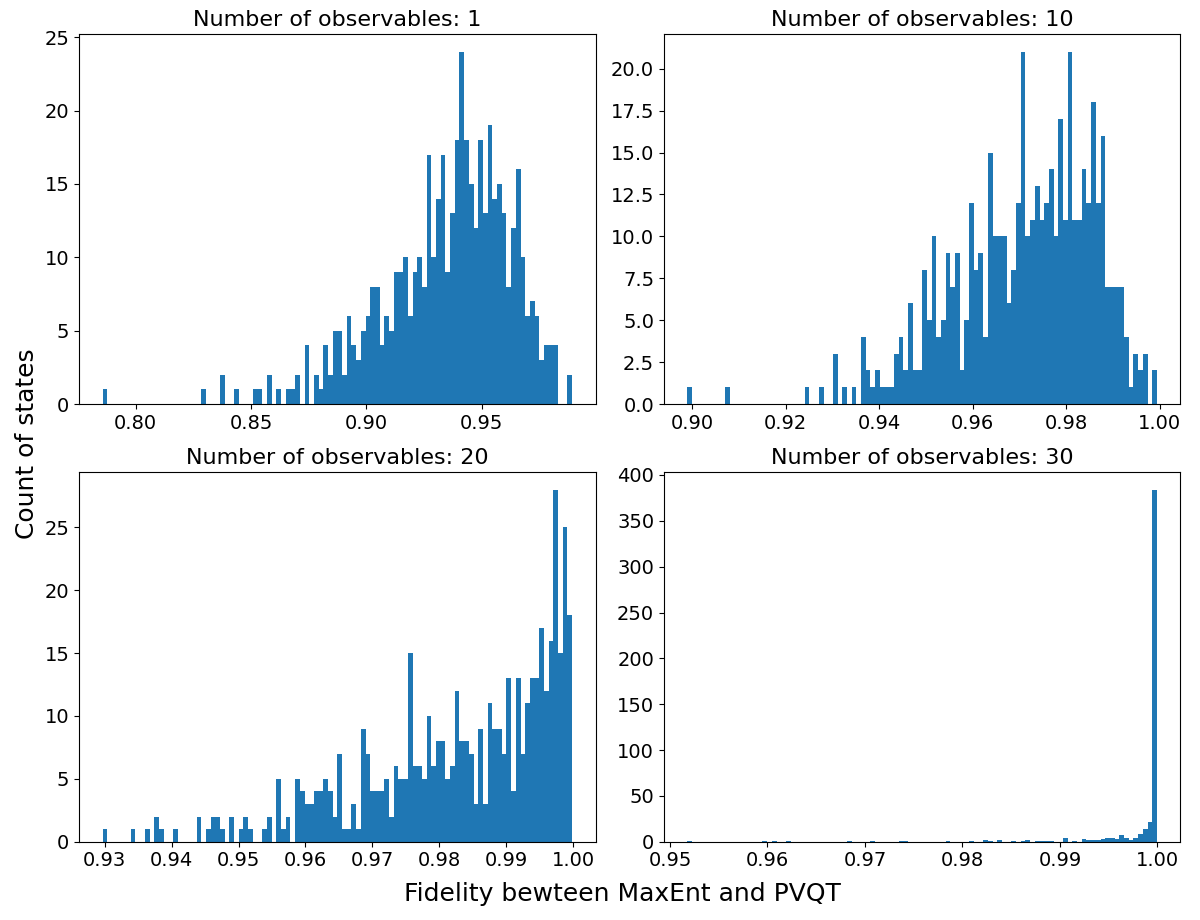}
      %
    \end{subfigure}\vspace{-10pt} 
    \hfill
    \begin{subfigure}{0.43\textwidth}
        \centering
            \caption{}
        \includegraphics[width=\linewidth]{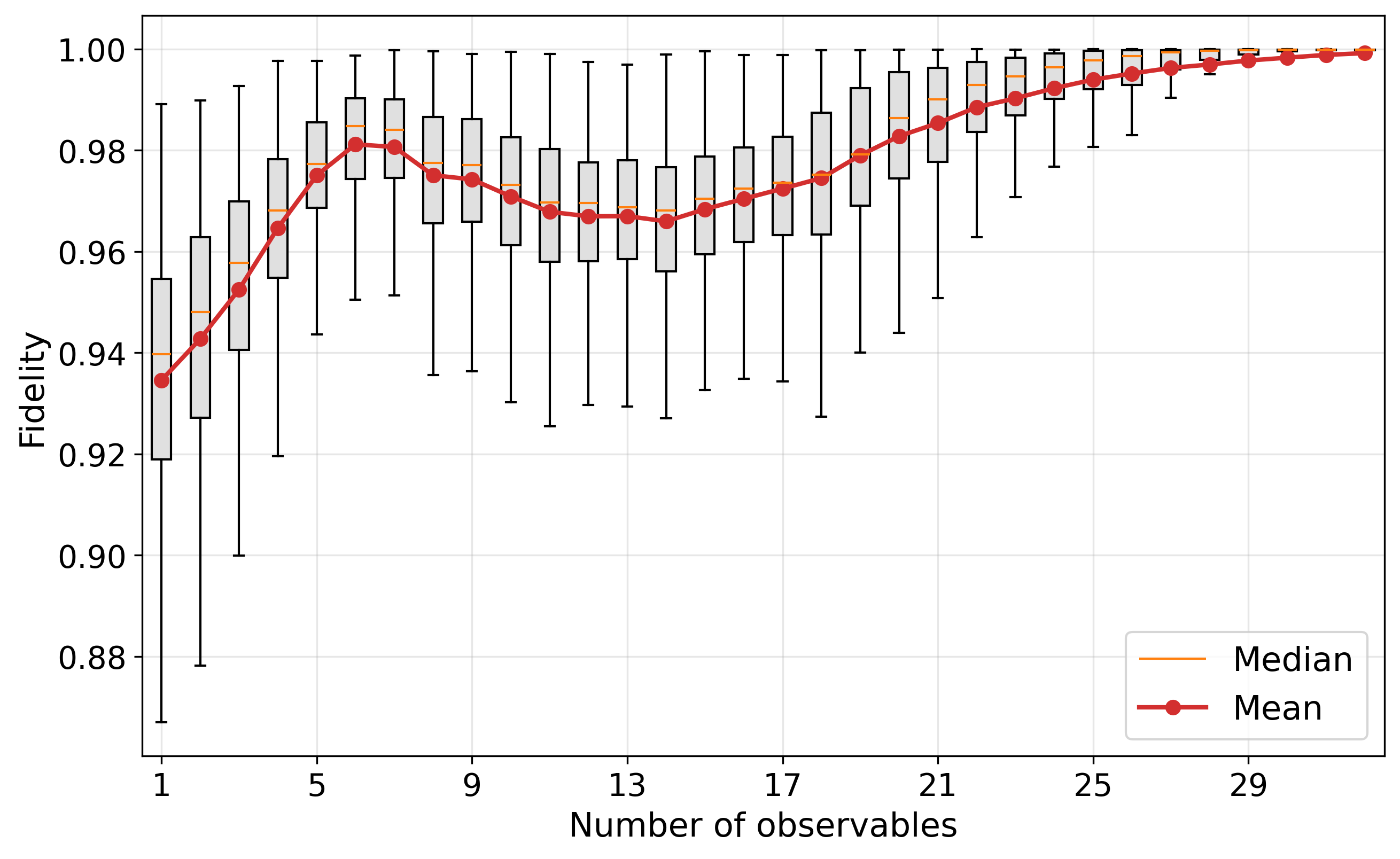}
%
    \end{subfigure}
    \vspace{-9pt}
    \caption{(a) 
    Histograms of fidelities between PVQT with  $\beta = 0.01$ and MaxEnt  reconstructed states, over $500$ random three-qubit pure states. 
    (b) 
    Average fidelity between PVQT with $\beta= 0.01$ and MaxEnt reconstructed states. Error bars represent quartiles. Note that mean (average) and median are close to each other but they are not equal.
    }
    \label{fig:hist-3q-and-quart-vqt-hib}
\end{figure}

As before, we compute histograms of the fidelity between the MaxEnt and PVQT reconstructed state, shown in Fig.~\ref{fig:hist-3q-and-quart-vqt-hib}a.
The histograms for PVQT with $\beta = 0.01$ are slightly shifted toward higher fidelities compared to those in Fig.~\ref{fig:hist-3q-and-quart-vqt-inf}a.
This can also be seen in the 
in the dispersion plots of Fig.~\ref{fig:hist-3q-and-quart-vqt-hib}b where the distributions are shifted to higher fidelity values compared to Fig.~\ref{fig:hist-3q-and-quart-vqt-inf}b. 
Furthermore, as the number of observables increases, the fidelity between the PVQT and MaxEnt reconstructed states approaches $1.0$, consistent with the previous results.

\begin{figure}
\centering
\includegraphics[width=8.8cm]{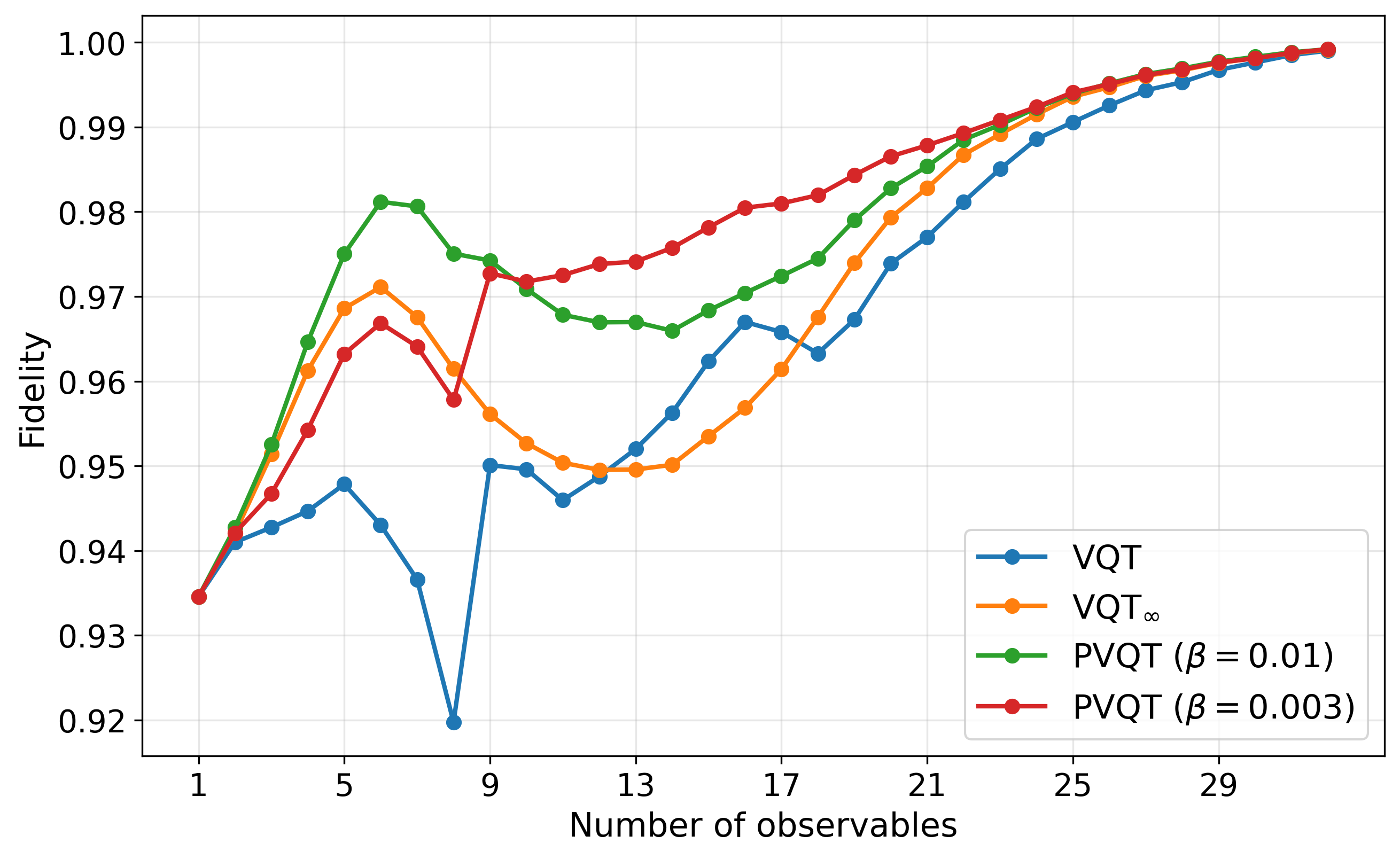}
\caption{Average fidelity between different VQTs and MaxEnt reconstructed states for $500$ random three-qubit pure states as a function of the number of measured observables.}
\label{vqts-vs-maxent-3q}
\end{figure}

It is worth noting that a higher fidelity with respect to the target state does not necessarily imply closer agreement with the MaxEnt reconstructed state, 
as shown in Fig.~\ref{vqts-vs-maxent-3q}.
In this figure we show the average fidelity between different VQTs and MaxEnt reconstructed states for $500$ random three-qubit pure states as a function of the number of measured observables.
It reveals an interesting behavior: the $\beta$ parameter can improve the fidelity between the PVQT and MaxEnt reconstructed states relative to that between VQT$_{\infty}$ and MaxEnt, for a certain number of observables.
Nevertheless, this effect does not happen uniformly across all ranges of $\beta$. In the case of $\beta=0.01$, the fidelity obtained with PVQT exceeds that of VQT$_{\infty}$ for all numbers of observables considered. In the $\beta=0.003$ case, PVQT outperforms VQT$_{\infty}$ only when the number of observables exceeds nine. Moreover, above this number, PVQT with $\beta=0.003$ yields higher fidelity than PVQT with $\beta=0.01$. In principle, one could select $\beta$ as a function of the number of observables to optimize the fidelity with respect to the ~MaxEnt state. 

\begin{figure}
\centering
\includegraphics[width=8.8cm]{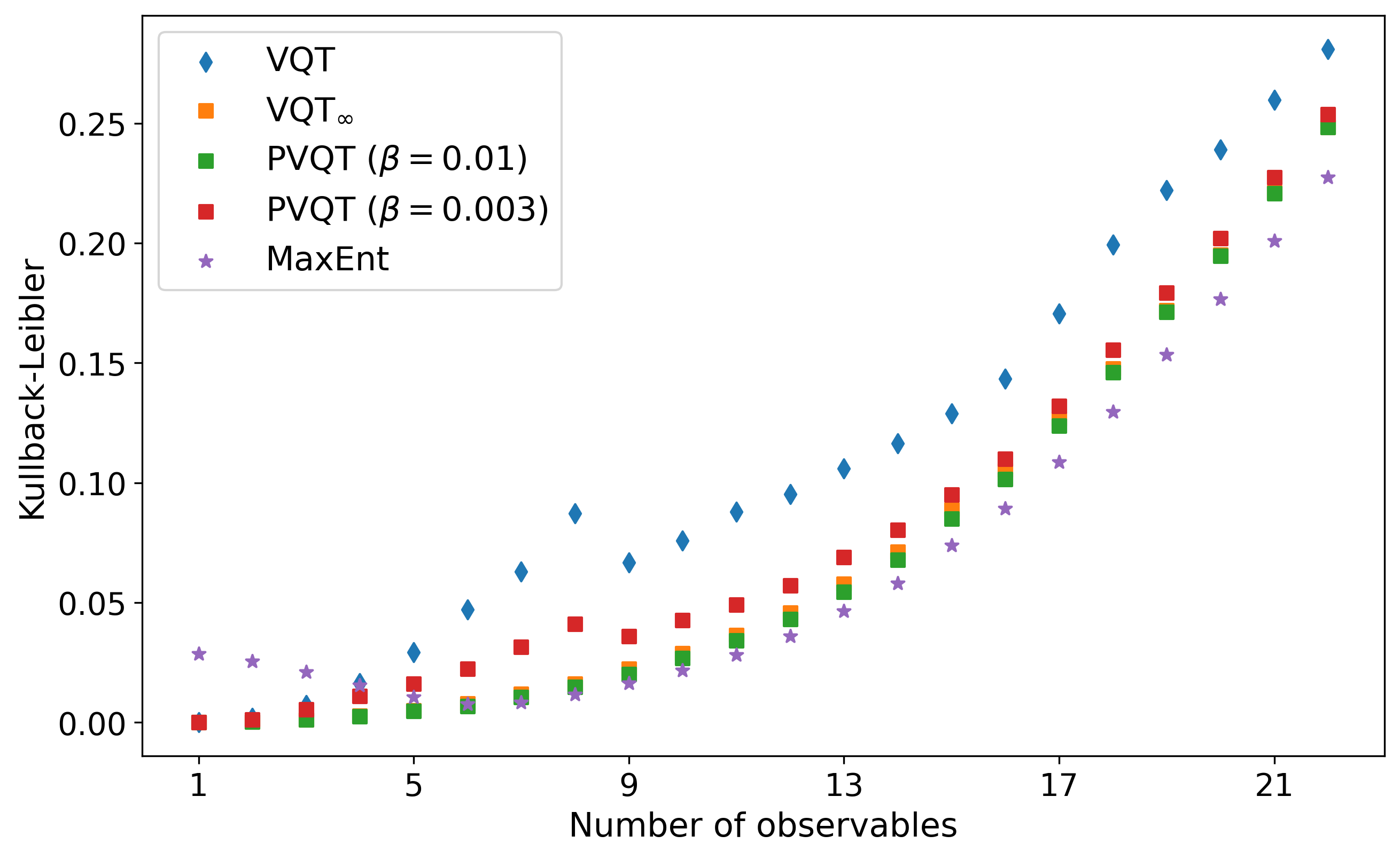}
\caption{
Average Kullback-Leibler divergence between the uniform distribution  and the unmeasured probability vector $\vec{u}$ of different VQTs and MaxEnt reconstructed states, for $500$ random three-qubit pure states, as a function of the number of measured observables. }
\label{fig:kl-3q}
\end{figure}

\begin{figure}
\centering
\includegraphics[width=8.8cm]{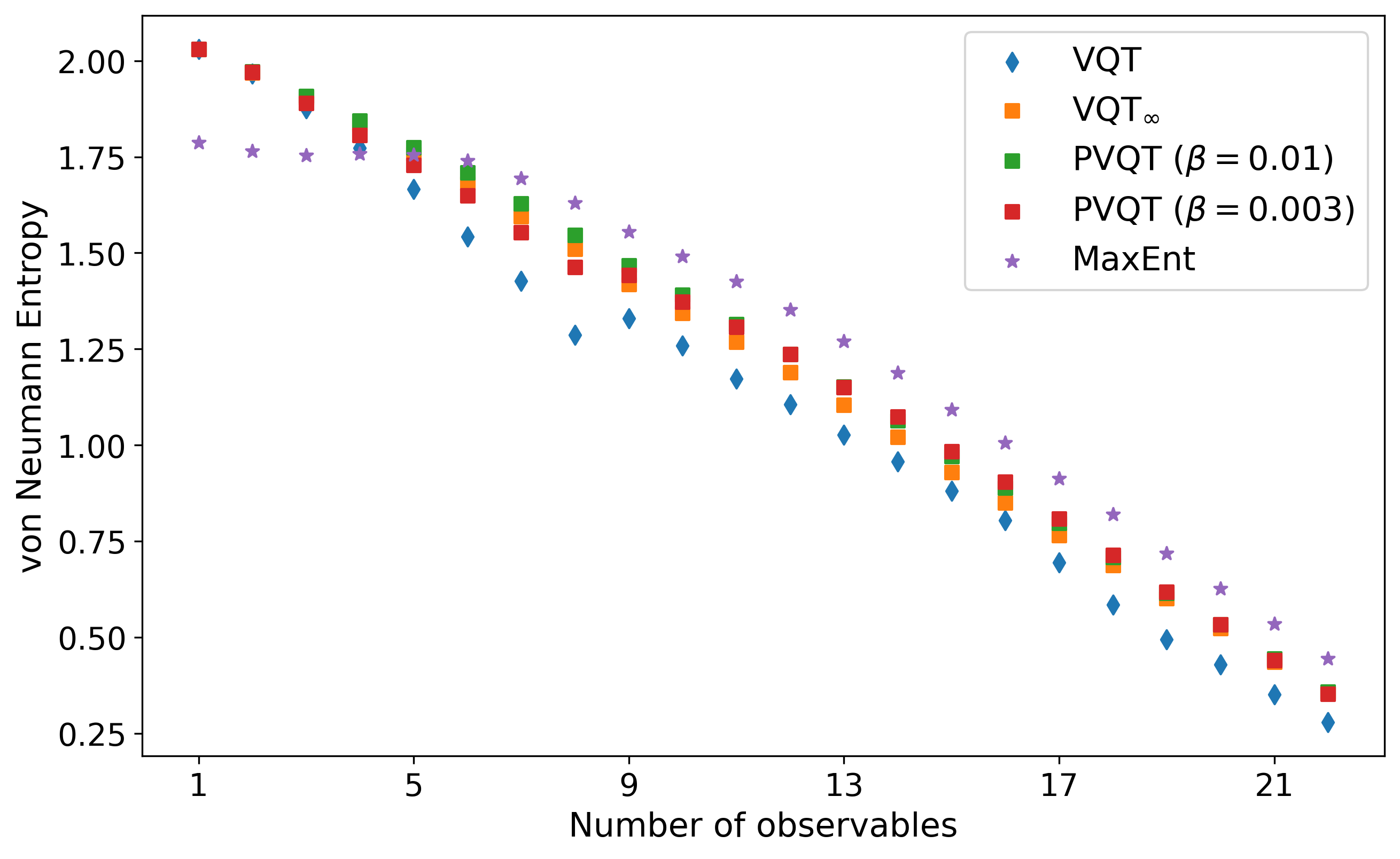}
\caption{
Average Von Neumann entropy for different VQTs and MaxEnt reconstructed states for $500$ random three-qubit pure states as a function of the number of measured observables. 
}
\label{fig:vnentropy-3q}
\end{figure}

To further assess the performance of PVQT, we compute the Kullback-Leibler divergence between the uniform distribution and  the unmeasured probability vector $\vec{u}$   of reconstructed states, defined in Eq. \eqref{vec_u}.
In \cite{Goncalves2013}, it was shown that in the eigenbasis case, MaxEnt method assigns a uniform distribution for this vector. 
Furthermore, it was numerically shown that when measuring SIC-POVM 
observables, the vector $\vec{u}$ obtained via VQT$_{\infty}$ and MaxEnt is closer 
to the uniform distribution than that obtained via VQT.
Fig.~\ref{fig:kl-3q} shows, however, that the unmeasured probability vector 
associated with PVQT for $\beta=0.003$ is less uniform than that of VQT$_{\infty}$. 
This is expected, since PVQT with $\beta=0.003$ approaches VQT. Despite this, 
PVQT with $\beta=0.003$ yields a higher von Neumann entropy than both VQT$_{\infty}$ 
and VQT when the number of observables exceeds nine, as shown in 
Fig.~\ref{fig:vnentropy-3q}, consistent with the higher fidelity of PVQT relative 
to VQT$_{\infty}$ observed in Fig.~\ref{vqts-vs-maxent-3q}.

These results indicate that there exist states that can be closer to the ~MaxEnt solution than those obtained via VQT$_{\infty}$, even when their unmeasured probability vectors are not as close to the uniform distribution as those of VQT$_\infty$. Consequently, characterizing proximity to ~MaxEnt solely through the ~Kullback–Leibler divergence of $\vec{u}$ with respect to the uniform distribution may fail to capture this behavior.


As a final comparison, we present results for higher-rank states with uniform noise, showing that the behavior reported in the previous section persists. Following \cite{Goncalves2013}, we plot 
average fidelity between different VQTs and MaxEnt reconstructed states as a function of the number of measured observables,
with 5$\%$ uniform noise added to the measurement outcomes. That means each outcome is perturbed by a value drawn uniformly from an interval of 5$\%$ around the true probability.
We use Eq. \eqref{eq:nlsq} for the MaxEnt method with noisy data.
 
\begin{figure}[H]
\centering
\includegraphics[width=8.8cm]{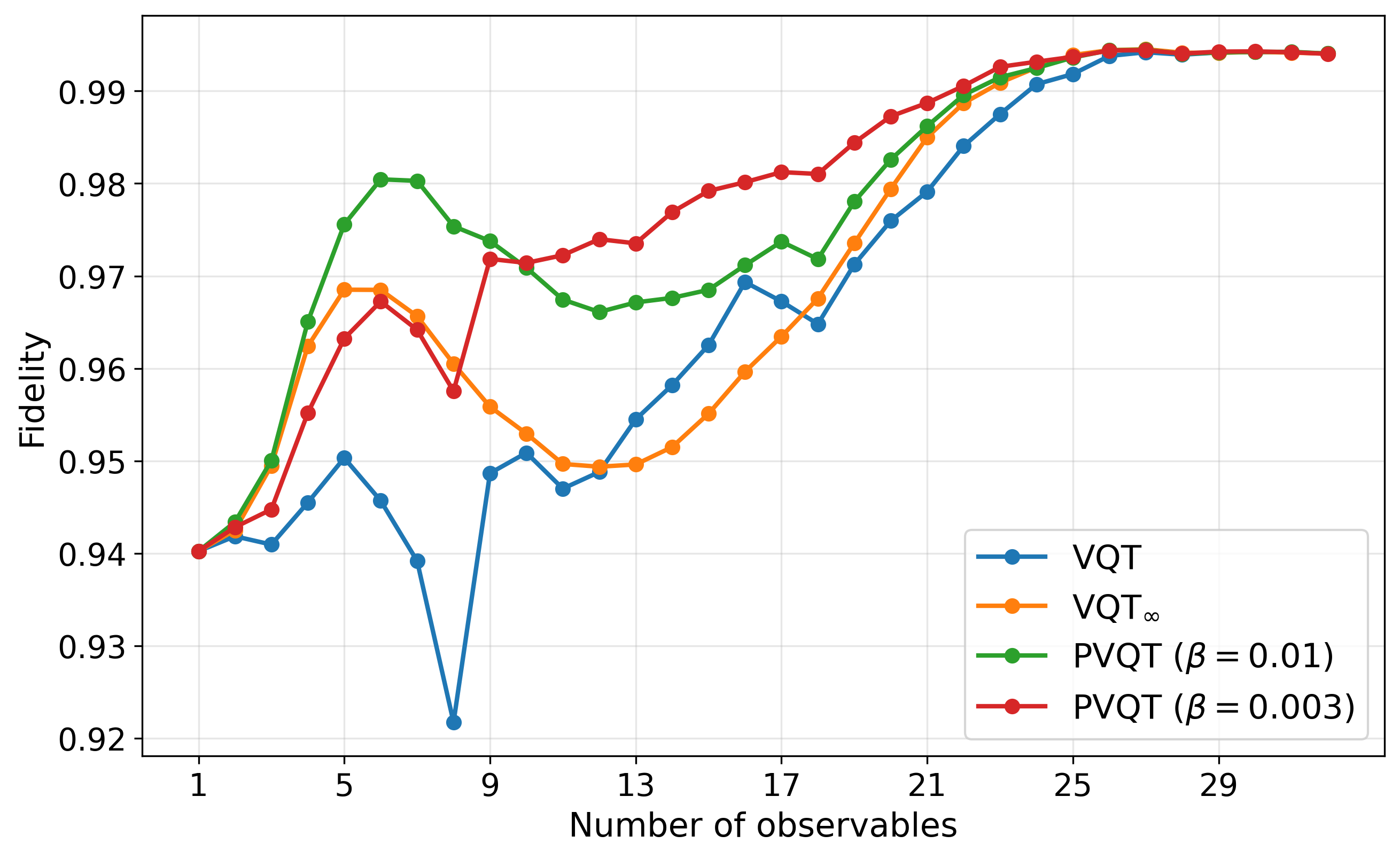}
\caption{Average fidelity between different VQTs and MaxEnt reconstructed states for $500$ random three-qubit pure states as a function of the number of measured observables, with 5$\%$ uniform noise added to the measurement outcomes.}
\label{vqts-vs-maxent-ruido-uniforme-3q}
\end{figure}

\begin{figure}[H]
\centering
\includegraphics[width=8.8cm]{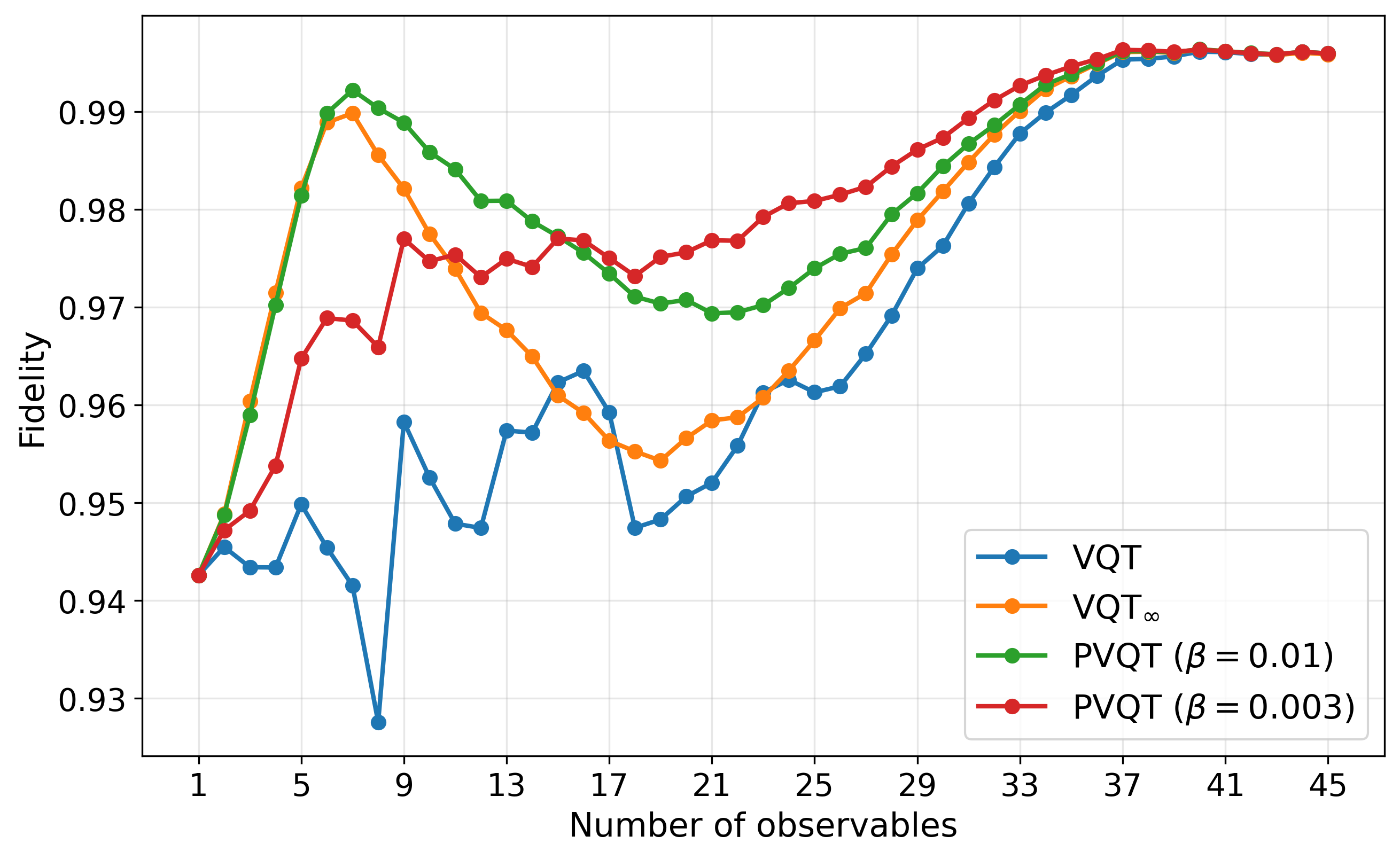}
\caption{
Average fidelity between different VQTs and MaxEnt reconstructed states for $500$ random three-qubit rank-$2$ states as a function of the number of measured observables, with $5\%$ uniform noise added to the measurement outcomes.
Cutoff around 45 observables.}
\label{vqts-vs-maxent-rango2-con-ruido-3q}
\end{figure}

Fig.~\ref{vqts-vs-maxent-ruido-uniforme-3q} and \ref{vqts-vs-maxent-rango2-con-ruido-3q} show 
average fidelities between different VQTs and MaxEnt reconstructed states for three-qubit states of rank one and two, respectively.

For rank-1 states, Fig.~\ref{vqts-vs-maxent-ruido-uniforme-3q}, a slight deviation from the curves shown in Fig.~\ref{vqts-vs-maxent-3q} is observed for a small number of observables. For rank-2 states, Fig.~\ref{vqts-vs-maxent-rango2-con-ruido-3q}, a larger number of observables is needed to reach a trace distance of $10^{-4}$  (around 45 observables), in agreement with the observation in \cite{Goncalves2013} that higher-rank states require more observables to achieve higher fidelity.

In both Figs.~\ref{vqts-vs-maxent-ruido-uniforme-3q} and 
\ref{vqts-vs-maxent-rango2-con-ruido-3q}, PVQT achieves better agreement with 
MaxEnt than VQT$_{\infty}$ for a suitable choice of $\beta$.

\subsection{Four- and five-qubit states}\label{subsec:4-5q}
As in the three-qubit case, the results for four- and five-qubit states exhibit the same trend. 
For the four-qubit case, we sampled 100 random pure states and set a trace distance cutoff of $10^{-3}$, which corresponds to around 65 observables. Note that the quorum for four qubits is 256.

\begin{figure}[H]
\centering
\includegraphics[width=8.8cm]{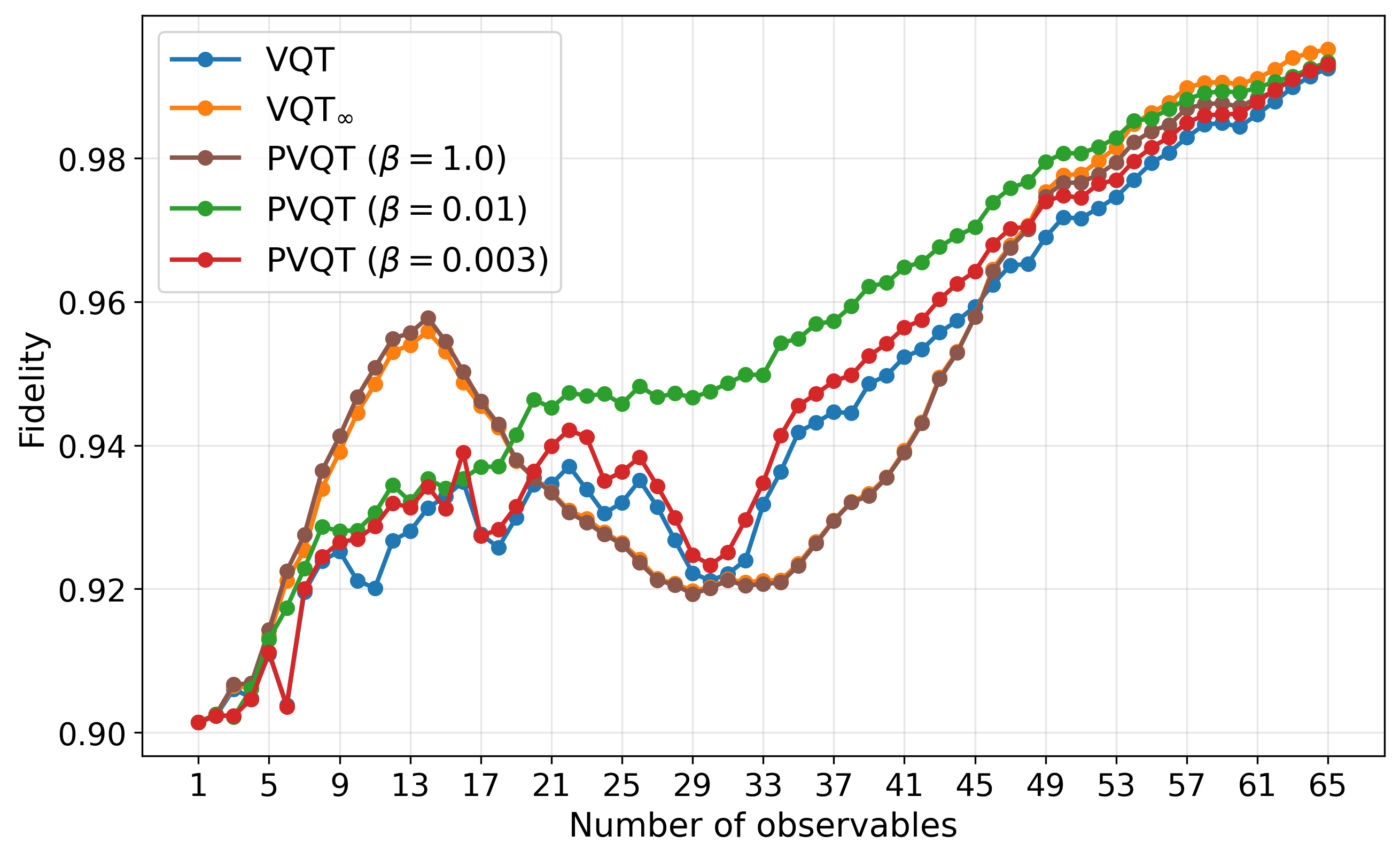}
\caption{Average fidelity between different VQTs and MaxEnt reconstructed states for $100$ random four-qubit pure states as a function of the number of measured observables.}
\label{vqts-vs-maxent-4q}
\end{figure}

In Fig.~\ref{vqts-vs-maxent-4q}, we show average fidelities between MaxEnt and different VQTs  reconstructed states as a function of the number of measured observables. 
As in the three-qubit case, VQT performs worse than 
VQT$_{\infty}$. Here, $\beta=0.01$ improves fidelity only beyond 18
observables. As expected, $\beta=0.003$ drives PVQT toward VQT, while $\beta=1$ 
drives it toward VQT$_{\infty}$. In this four-qubit case, one can treat $\beta$ 
as a function of the number of observables, yielding higher fidelity than 
VQT$_{\infty}$.

\begin{figure}[H]
\centering
\includegraphics[width=8.8cm]{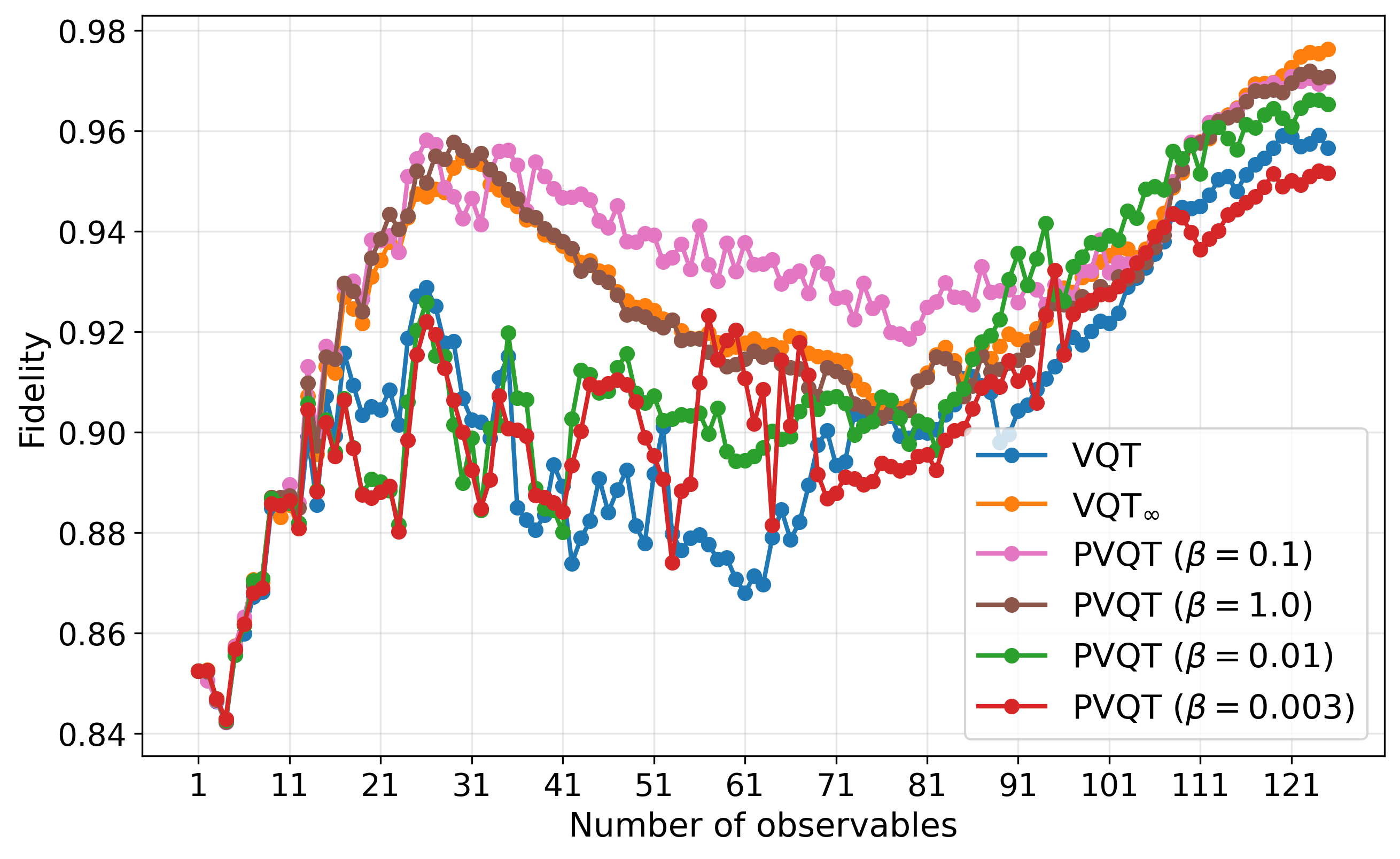}
\caption{Average fidelity between different VQTs and MaxEnt reconstructed states for $10$ random five-qubit pure states as a function of the number of measured observables.}
\label{vqts-vs-maxent-5q}
\end{figure}

For the five-qubit case (Fig.~\ref{vqts-vs-maxent-5q}), the analysis is restricted 
to a small sample of 10 random pure states. Although this sample size is not 
sufficient for a statistically robust analysis, the results nonetheless reflect 
the same qualitative trend: $\beta$ can be adjusted to improve performance. 
Among the values of $\beta$ explored, PVQT with $\beta=0.1$ outperforms 
VQT$_{\infty}$ for approximately 31 to 100 observables.

\section{Conclusions}\label{concl}

In this work, we presented a comparative numerical analysis of VQT$_{\infty}$, 
MaxEnt, and PVQT. Using fidelity as a similarity metric, our results suggest 
that these methods yield similar reconstructed states on average, consistent 
with the findings of \cite{Goncalves2013} and further supported by our 
extended analysis.
In general POVM scenarios, the methods are not strictly equivalent, yet their reconstructed states remain in close agreement, as quantified by the fidelity.

Since the MaxEnt estimator is uniquely defined as the least biased inference 
compatible with the observed data, any deviation from it by VQT$_{\infty}$ may 
indicate the presence of undesirable bias. By reintroducing into the cost function 
a term proportional to the $1$-norm of the unmeasured probability vector and 
appropriately tuning the associated hyperparameters, this bias can be mitigated, 
as demonstrated by our numerical results.

Our simulations for systems of up to five qubits show that suitable optimization 
of the PVQT hyperparameters yields reconstructed states with higher fidelity and 
higher von Neumann entropy than those obtained with VQT$_{\infty}$. We also found 
that the Kullback-Leibler divergence does not fully capture this improvement, as 
it fails to reflect the higher von Neumann entropy of the PVQT reconstructed 
states relative to those of VQT$_{\infty}$. These improvements arise well below 
the number of observables required for informational completeness. Beyond a 
certain number of measured observables, all methods converge, identifying a regime 
in which they may be used interchangeably. This provides practical flexibility, allowing one to leverage the lower computational cost of VQT-based approaches without compromising agreement with MaxEnt reconstructed states.

Future work will address both numerical and theoretical challenges, including the 
systematic selection of hyperparameters, the simultaneous variation of $\alpha$ 
and $\beta$, and the extension of these methods to larger numbers of qubits.

\section*{Aknowledgements}
V.P. gratefully acknowledges the Cluster Group at CITECCA–UNRN for providing access to their computational resources.

\bibliographystyle{apsrev4-1}
\bibliography{main}

\end{document}